\documentclass[12pt]{article}
\pdfoutput=1
\usepackage{subfigure}
\usepackage{amssymb,amsmath}
\usepackage{graphicx}
\usepackage{color}
\usepackage[colorlinks=true
,urlcolor=blue
,citecolor=blue
,linkcolor=blue
,pagecolor=blue
,linktocpage=true
,pdfproducer=medialab
]{hyperref}
\usepackage[a4paper]{geometry}

\usepackage{placeins}

\makeatletter \renewcommand{\@dotsep}{10000} \makeatother

\mathchardef\mhyphen="2D

\newcommand{\beq}{\begin{equation}}
\newcommand{\eeq}{\end{equation}}
\newcommand{\bea}{\begin{eqnarray}}
\newcommand{\eea}{\end{eqnarray}}

\usepackage[nodisplayskipstretch]{setspace}

\begin{document}

\begin{titlepage}
\pagestyle{empty}

\begin{flushright}
FTPI-MINN-16/24
\end{flushright}

\vspace*{0.2in}
\begin{center}
{\Large \bf    Gauge Mediation Models with Adjoint Messengers
  }\\
\vspace{1cm}
{\bf  Ilia Gogoladze$^{a,}$\footnote{E-mail: ilia@bartol.udel.edu;\\
\hspace*{0.5cm} On leave of absence from Andronikashvili Institute
of Physics,  Tbilisi, Georgia.},
Azar Mustafayev$^{b,}$\footnote{E-mail: mustafayev@physics.umn.edu},
Qaisar Shafi$^{a,}$\footnote{E-mail: shafi@bartol.udel.edu} and
Cem Salih $\ddot{\rm U}$n$^{c,d,}\hspace{0.05cm}$\footnote{E-mail: cemsalihun@uludag.edu.tr}}
\vspace{0.5cm}

{\it
$^a$Bartol Research Institute, Department of Physics and Astronomy,\\
University of Delaware, Newark, DE 19716, USA \\
$^b$William I.~Fine Theoretical Physics Institute, \\ University of Minnesota, Minneapolis, MN 55455, USA\\
$^c$Department of Physics, Uluda\~{g} University, TR16059 Bursa, Turkey \\
$^d$ Center of Fundamental Physics, Zewail City of Science and Technology, 6 October City, 12588, Cairo, Egypt
}

\end{center}

\vspace{0.5cm}
\begin{abstract}
\noindent

We present a class of models in the framework of gauge mediation supersymmetry
breaking where the messenger fields transform in the adjoint representation of the Standard Model gauge symmetry.
To avoid unacceptably light right-handed sleptons in the spectrum 
we introduce a non-zero $U(1)_{B-L}$ D-term. 
This  leads to  an additional  contribution  to  the  soft  supersymmetry  breaking  mass  terms which
makes the right-handed slepton masses compatible with the current experimental bounds.
We show that in this framework the observed 125~GeV Higgs boson mass can be accommodated  
with the sleptons accessible at the LHC, while the squarks and gluinos lie in the multi-TeV range. 
We also discuss the issue of the fine-tuning and show that the
desired relic dark matter abundance can also be accommodated.

\end{abstract}

\end{titlepage}



\section{Introduction}
\label{ch:introduction}

Models with gauge mediated supersymmetry breaking (GMSB) provide a compelling resolution to the  supersymmetry (SUSY)
flavor problem since the soft SUSY breaking (SSB) terms are generated by
the flavor blind gauge interactions~\cite{Dine:1993yw,Giudice:1998bp,Meade:2008wd}.
In general, the trilinear SSB A-terms in GMSB scenarios are relatively small  at the messenger scale,
even if an additional sector is added to generate the $\mu/B\mu$ terms~\cite{Komargodski:2008ax}.
Because of the small A-terms, accommodating the light CP-even Higgs boson mass around 125~GeV requires a stop mass
in the  multi-TeV range~\cite{Ajaib:2012vc}. On the other hand, a multi-TeV top squark has a very strong influence
on the sparticle spectrum ~\cite{Ajaib:2012vc},  if we assume that the messenger fields reside in the SU(5) representations
such as  $5+\bar5$ or $10+\overline{10}$. This case is called minimal GMSB scenario, since it  is the simplest
scenario that preserves gauge coupling unification of the  minimal supersymmetric standard model (MSSM)
and  provides non-zero SSB mass terms for all supersymmetric particles.
It is often assumed that all  messenger fields have a universal mass in this simplest model.
Even if one assumes a large mass splitting among the colored and non-colored messenger fields,
the sparticle mass spectrum cannot be entirely separated, since all  fields  from $5+\bar5$
(or $10+\overline{10})$  representation have non-zero hypercharge, and they can generate non-zero masses
through hypercharge interactions. This means that in these models the maximal splitting among sfermion SSB mass terms
cannot exceed the ratio  of corresponding fermion hypercharges.
Therefore, the colored and non colored sparticle mass spectra are closely linked here.
For instance, if we have a multi-TeV mass  top squark in the minimal GMSB scenario,
then the whole SUSY sparticle spectrum is also around the TeV scale~\cite{Ajaib:2012vc}.
Note that  $t$-$b$-$\tau$ Yukawa coupling unification can be realized in these models and it provides
a specific spectrum for sparticle masses~\cite{Gogoladze:2015tfa}.

Sometime ago it was {proposed}~\cite{Han:1998pa} that the messenger fields should reside in the adjoint representations of  $SU(3)_C \times SU(2)_L$, namely in
 $[(8,1) + (1,3)]$. In this scenario the colored and non-colored sparticle mass spectra can be significantly separated from each other
since these  multiplets do not carry hypercharge, and so there is no common contribution to SSB mass terms from $U(1)_Y$ interaction.
In this case we can  have multi-TeV stop masses, while the sleptons potentially can be much lighter $\gtrsim O(100)$~GeV. Unfortunately, in this scenario  the right-handed sleptons and bino do not obtain
 their SSB mass terms at the same loop level as other sparticles do, the reason being that the right-handed sleptons and bino do not transform
 under  $SU(3)_C \times SU(2)_L$ and the messenger fields in  $[(8,1) + (1,3)]$  do not have hypercharge.
 This is in disagreement, potentially, with the experimental lower bound of $100$ GeV on the charged slepton masses~\cite{Agashe:2014kda}.
 In order to generate a mass for right-handed sleptons of $O(100)$~GeV through renormalization group equation (RGE) running,
 the some other appropriate sparticle masses need to be in the multi-TeV range.  To solve this problem, an additional
 source which contributes to the SSB mass terms 
 can be proposed. For instance, a messenger field from $(5+\bar5)$ can be included~\cite{Han:1998pa}, in addition to those in $[(8,1) + (1,3)]$.
 Another proposal is for the bino mass to be generated through the gravitational interactions~\cite{Bhattacharyya:2013xba}. The bino mass is effective in generating a SSB mass term for the right-handed slepton through RGE running~\cite{Martin:1997ns}, and
 it was simply assumed that a universal SSB mass term~\cite{Bhattacharyya:2015vha}  be added to the sparticle masses generated from  $[(8,1) + (1,3)]$ messenger fields, when the bino is very light.

In general, D-term contribution to the scalar masses can arise whenever a gauge symmetry is spontaneously broken with reduction of rank~\cite{Drees:1986vd}.
Here we propose a scenario where  all MSSM sfermions obtain additional SSB mass terms from a non-zero $U(1)_{B-L}$ $ D$-term~\cite{Kolda:1995iw}.
$U(1)_{B-L}$~\cite{Mohapatra:1980qe} is one of the most natural extensions of the SM gauge symmetry,  and it is also
part of $SO(10)$ grand unified theory~\cite{georgi} or Pati-Salam model~\cite{pati}, which are considered to be compelling extensions of the SM.
In our scenario, the right-handed sleptons obtain their masses only from the $U(1)_{B-L}$  $D-$term contribution. The bino mass vanishes at the messenger scale
 and is generated at low scale through RGE evolution. As we will show, in our scenario the bino mass can lie in the $O$(GeV)$-O$(100\,\rm~GeV) interval.

It has been shown in Ref.~\cite{Gogoladze:2009bd}  that non-universal gaugino masses at  the GUT scale ($M_{\rm GUT}$)
 can help resolve the little hierarchy problem in gravity mediated scenario~\cite{Chamseddine:1982jx}.  In particular,
 the little hierarchy problem  can be resolved if the ratio between $SU(2)_L$ and $SU(3)_c$
gaugino masses satisfies the asymptotic relation $M_2/M_3\approx 3$~\cite{Gogoladze:2009bd}.
In this case the leading contributions to  $m^{2}_{H_{u}}$ through  RGE  evolution  are proportional to $M_2$ and $M_3$ and can cancel each other.
This allows for large  values of  $M_2$ and $M_3$ in the gravity mediated supersymmetry  breaking scenario~\cite{Chamseddine:1982jx},
while keeping the value of $m^{2}_{H_{u}}$ relatively small. On the other hand, large values of $M_2$ and $M_3$ yield a heavy top squark ($>$ TeV),
which is necessary to accommodate   $m_h \simeq 125$~GeV. A similar observation was made in GMSB scenario with non-universal gaugino masses at the messenger scale~\cite{Brummer:2012zc}.
We also obtain in our scenario a relatively light MSSM $\mu$-term which helps  ameliorate the little hierarchy problem
at the electroweak scale $(M_{\rm EW})$~\cite{Baer:2012mv}.

The remainder of this paper is organized as follows:  We present the model  in Section~\ref{sec:model} and in Section~\ref{sec:constraints}
we summarize the scanning procedure and the experimental constraints we employ.
In Section~\ref{sec:results} we present our results focusing on the 
low mass spectrum for the sleptons and accommodating the 125~GeV Higgs boson mass
and relic dark matter abundance. We also provide in this section a table of benchmark points which exemplifies our findings.
Our conclusions are discussed in Section~\ref{sec:conclusion}.
%

\section{Essential Features of the Model}
\label{sec:model}
%

Supersymmetry breaking in a typical GMSB scenario takes place in a hidden sector, and this breaking is transferred to the visible sector via messenger fields.
These messenger fields interact with the visible sector via the SM gauge interactions and induce the SSB terms in the MSSM through loops.
In order to preserve perturbative gauge coupling unification, the minimal GMSB  scenario  can include $N_5$  pairs of $(5+\overline{5})$  ($N_{5}=1, ... , 5$)
or one $(10+\overline{10})$ pair, or one combination $10+\overline{10}+5 +\overline{5}$, or  one pair of $15+\overline{15}$ of $SU(5)$ multiplets~\cite{Giudice:1998bp}.
On the other hand, it was proposed in~\cite{Han:1998pa}  to have the messenger fields reside in $[(8,1,0) + (1,3,0)]$  representations  of
$SU(3)_C \times SU(2)_L \times U(1)_Y$. In this scenario the colored and non-colored sparticle spectra can be significantly separated
from each other, the reason being that the messenger fields  do not carry hypercharge, and so there is no common contribution to the SSB mass terms from $U(1)_Y$ interaction.
This  allows one to have relatively light sleptons and electroweak gauginos which can be accessible at the LHC,
while the gluino and stop can be in the multi-TeV mass range.

In this paper we will study the scenario in which the fields in $[(8,1,0) + (1,3,0)]$ are the messengers of SUSY breaking.
We also propose an additional contribution to the SSB masses of the sfermions from the $D-$term associated with $U(1)_{B-L}$ gauge group
to avoid inconsistently light right-handed slepton solutions. The bino in our scenario, as we will show, obtains a sizable mass through RGE evolution.
In order to incorporate SUSY breaking in the messenger sector, the fields in $[(8,1,0)(\Sigma_8) + (1,3,0)(\Sigma_3)]$ dimensional multiplets are coupled,
say, with the hidden sector gauge singlet chiral field $S$~\cite{Han:1998pa},
\begin{equation}
 W \supset (m_3+\lambda_3\, S) {\rm Tr}(\Sigma_3^2)  + (m_8+\lambda_8\, S) {\rm Tr} (\Sigma_8^2).
\end{equation}
Here, for simplicity, we assume $M_{\mathrm{Mess}}\equiv m_3=m_8$, and  the $F_S$ component of $S$ has a non-zero vacuum expectation value (VEV).
$W$ denotes the appropriate superpotential of the model.
Below the messenger scale $M_{\mathrm{Mess}}$, the fields $\Sigma_3$ and $\Sigma_8$ decouple generating SSB masses for the MSSM fields.
The  gaugino masses generated at  one-loop level  are given by
\beq
M_{1}=0 ~,\hspace{0.3cm}M_{2}\simeq \frac{g_{2}^{2}}{16\pi^{2}}2\Lambda_{3}~\hspace{0.3cm}
M_{3}\simeq \frac{g_{3}^{2}}{16\pi^{2}}3\Lambda_{8}~,
\label{BCgaugino}
\eeq
 where $i = 1,\, 2,\, 3,$  stand for the  $U(1)_Y$,  $SU(2)_L$, and  $SU(3)_c$   sectors, respectively,
 and $\Lambda_3 = \lambda_3\langle F_S \rangle / M_{\mathrm{Mess}}$ and  $\Lambda_8 = \lambda_8\langle F_S \rangle / M_{\mathrm{Mess}}$.
 The bino mass $M_1$ will be generated at the two-loop level~\cite{Yamada:1993ga}, and  it vanishes at the messenger scale.
 As we will show, the RGE evolution with the relevant SUSY  parameters results in bino masses of around 100~GeV or so.

The SSB  masses for the MSSM scalars induced at  two-loop level are as follows~\cite{Han:1998pa}
\begin{eqnarray}
&& m^{2}_{\tilde{Q}} \simeq \dfrac{2}{(16\pi^{2})^{2}}\left[ \dfrac{4}{3}g_{3}^{4}3\Lambda^{2}_{8}+\dfrac{3}{4}g_{2}^{4}2\Lambda_{3}^{2} \right] \nonumber \\
&& m^{2}_{\tilde{U}} =  m^{2}_{\tilde{D}} \simeq  \dfrac{2}{(16\pi^{2})^{2}}\left[\dfrac{4}{3}g_{3}^{4}3\Lambda_{8}^{2} \right] \nonumber \\
&& m^{2}_{L} \simeq \dfrac{2}{(16\pi^{2})^{2}}\left[ \dfrac{3}{4}g_{2}^{4}2\Lambda_{3}^{2} \right] \nonumber \\
&& m^{2}_{H_{u}} =  m_{H_{d}}^{2} = m^{2}_{L} \nonumber \\
&& m_{E}^{2} =  0~.
\label{BCscalar}
\end{eqnarray}
The right-handed slepton masses will be generated at a higher loop level, and thus, they vanish at the messenger scale.
However, they are generated below the messenger scale from the RGE evolution. On the other hand, experiments require that the sleptons must be heavier than 100~GeV or so. In order to generate right-handed slepton masses of order 100~GeV or higher in this model,
some of the other sparticles should be around 100~TeV or so, which makes supersymmetry much less motivated for solving the gauge hierarchy problem.
To avoid this problem we consider an extension of  the SM gauge symmetry with $U(1)_{B-L}$. In this case it is natural to assume
that the D-term associated with $U(1)_{B-L}$ can provide a non-zero contribution~\cite{Gogoladze:2015jua} to the scalar SSB mass terms.
In summary,  the MSSM sfermion  masses have the following expression
\beq
  m_{\phi_{i}}^{2}=(m_{\phi_{i}}^{2})_{{\rm GMSB}}+e_{\eta}^{2}D^{2}~,
 \label{sGMSB_masses3}
\eeq
where  $\phi_{i}$  denote the MSSM sfermions, and $e_{\eta}$ stands for the sparticle charges under $U(1)_{B-L}$~\cite{Mohapatra:1980qe}.

The A-terms in our scenario  vanish  at the messenger scale, which is very common in GMSB models
(except when the MSSM and messenger fields are mixed~\cite{Evans:2015swa}, which we do not consider in this study).
The A-terms, as usual,  are generated from the RGE running and are small compared to the top squark mass.
The bilinear SSB term also vanishes at $M_{\mathrm{Mess}}$, although it is often ignored. We do not impose the relation $B_{\mu}=0$,
anticipating that the value needed to achieve the electroweak symmetry breaking (EWSB)
can be explained by some suitable mechanisms operating at the messenger scale~\cite{Komargodski:2008ax}.

The sparticle spectrum in our model is therefore completely specified by the following parameters defined at the messenger scale:
\beq
M_{\mathrm{Mess}}, \, \Lambda_3, \,   \Lambda_8, \, \mathrm{tan}\beta,  \, sign({\mu}), \,  N_5,\,   c_{\rm grav}, \, D~,
\label{mgmsb-params1}
\eeq
where $M_{\mathrm{Mess}}$, $\Lambda_{8}$, and $\Lambda_{3}$ are defined earlier. ${\rm tan\beta}$ is the ratio of the VEVs of the two MSSM Higgs
doublets. The magnitude of $\mu$, but not its sign, is determined by the
radiative electroweak breaking (REWSB) condition. The parameter $c_{\rm grav} (\geq 1)$ affects the mass of the gravitino
and we set it equal to unity from now on. For simplicity, we consider the case $N_5=1$. Changing the  value of $N_5$ does not significantly alter
the sparticle spectrum~\cite{Ajaib:2012vc}. Finally, $D$ denotes the $D-$term contribution associated with $U(1)_{B-L}$.

Even though the messenger multiplets in this model are incomplete SU(5) multiplets, gauge coupling unification can still be achieved
 if we assume that the masses of $[(8,1,0) + (1,3,0)]$  fields are around $10^{13}$~GeV or so~\cite{Bachas:1995yt}.
 Note that in this case the gauge coupling unification scale is higher than the conventional SUSY GUT scale $\sim 2\times 10^{16}$~GeV.
In principle, it can be as high as the Planck scale.

\section{Scanning Procedure and Experimental Constraints}
\label{sec:constraints}

For our scan over the fundamental parameter space of GMSB with the adjoint messengers, we employed ISAJET 7.84 package~\cite{Paige:2003mg}
supplied with appropriate boundary conditions at $M_{\rm Mess}$. In this package, the weak-scale values of gauge and Yukawa
couplings are evolved from $M_Z$ to $M_{\mathrm{Mess}}$ via the MSSM RGEs in the $\overline{DR}$ regularization scheme.
For simplicity, we do not include the Dirac neutrino Yukawa coupling in the RGEs, whose contribution is expected to be small.

The SSB terms are induced at the messenger scale and we set them according to
Eqs.~(\ref{BCgaugino})-(\ref{sGMSB_masses3}).
From $M_{\mathrm{Mess}}$ the SSB parameters, along with the gauge and Yukawa couplings, are evolved down to the weak scale $M_Z$.
In the evolution of Yukawa couplings the SUSY threshold
corrections~\cite{Pierce:1996zz} are taken into account at the common scale $M_{{\rm SUSY}} = \sqrt{m_{\tilde{t}_{L}}m_{\tilde{t}_{R}}}$,
where $m_{\tilde{t}_{L}}$ and $m_{\tilde{t}_{R}}$ are the soft masses of the third generation left and right-handed top squarks respectively.

We have performed random scans over the model parameters given in Eq.~(\ref{mgmsb-params1}) in the following range:
\bea
\label{parameterRange}
10^{4}\leq & \Lambda_{3} & \leq 10^{6} {\rm GeV} \nonumber \\
10^{4}\leq & \Lambda_{8} & \leq 10^{6} {\rm GeV} \nonumber \\
10^{4}\leq & M_{{\rm Mess}} & \leq 10^{16} {\rm GeV} \\
0 \leq & D & \leq 2000 {\rm GeV} \nonumber \\
2 \leq & \tan\beta & \leq 60 \nonumber\\
N_5=1, & \mu > 0, & c_{\rm grav}=1~.  \nonumber
\eea

Regarding the MSSM parameter $\mu$, its magnitude but not the sign is determined by the radiative electroweak symmetry breaking (REWSB).
In our model we set ${\rm sign}(\mu)=1$. Finally, we employ the current central value for the top mass, $m_{t}=173.3$~GeV.
Our results are not too sensitive to one or two sigma variation of $m_{t}$~\cite{Gogoladze:2011db}.

In scanning the parameter space, we employ the Metropolis-Hastings
algorithm as described in Ref.~\cite{Belanger:2009ti}. The data points collected all satisfy
the requirement of radiative electroweak symmetry breaking (REWSB). We successively apply mass bounds
including the Higgs boson~\cite{:2012gu,:2012gk} and gluino masses~\cite{gluinoLHC},
and the constraints from the rare decay processes $B_s \rightarrow \mu^+ \mu^-$~\cite{BsMuMu},
$b \rightarrow s \gamma$~\cite{Amhis:2012bh} and $B_u\rightarrow\tau \nu_{\tau}$~\cite{Asner:2010qj}.
The constraints are summarized below in Table~\ref{table1}.
\begin{table}[h]\centering
\begin{tabular}{rlc}
$   123\, {\rm~GeV} \leq m_h \leq127$ \,{\rm~GeV} &
\\
$ m_{\tilde{g}} \geq 1.8$ \,{\rm TeV} & \\
$0.8\times 10^{-9} \leq{\rm BR}(B_s \rightarrow \mu^+ \mu^-)
  \leq 6.2 \times10^{-9} \;(2\sigma)$ &
\\
$2.99 \times 10^{-4} \leq
  {\rm BR}(b \rightarrow s \gamma)
  \leq 3.87 \times 10^{-4} \; (2\sigma)$ &
\\
$0.15 \leq \frac{
 {\rm BR}(B_u\rightarrow\tau \nu_{\tau})_{\rm MSSM}}
 {{\rm BR}(B_u\rightarrow \tau \nu_{\tau})_{\rm SM}}
        \leq 2.41 \; (3\sigma)$ & .
\end{tabular}
\caption{Phenomenological constraints implemented in our study.}
\label{table1}
\end{table}

\section{Results}
\label{sec:results}

\begin{figure}[t!]
\subfigure{\includegraphics[scale=1]{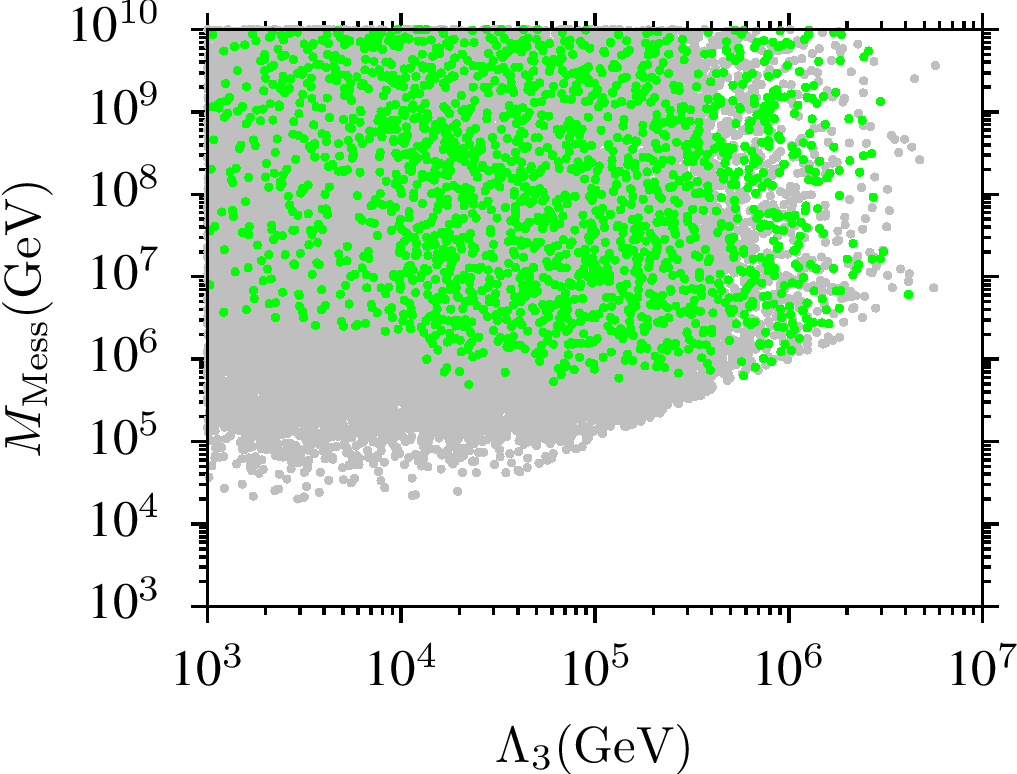}}
\subfigure{\includegraphics[scale=1]{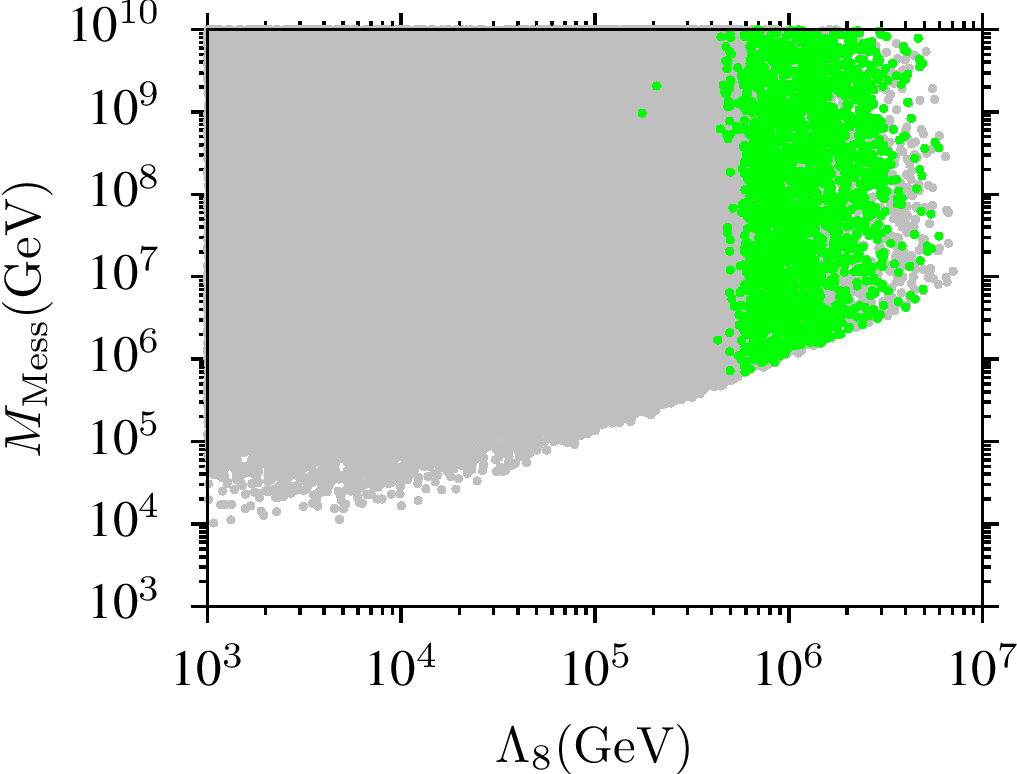}}
\subfigure{\includegraphics[scale=1]{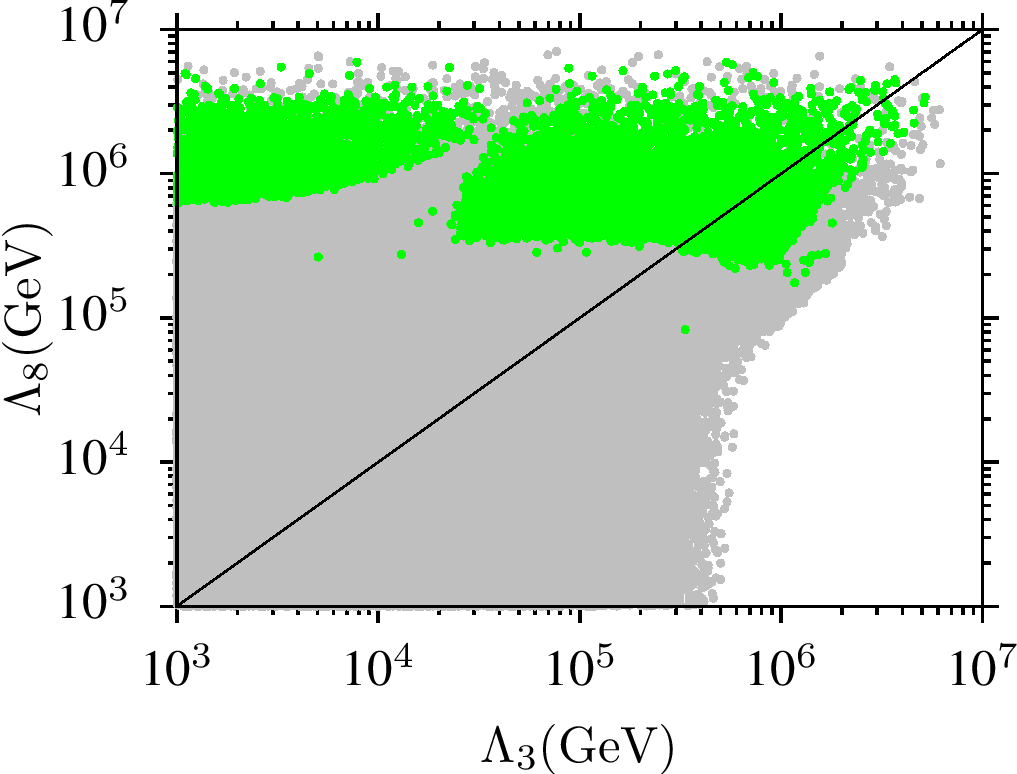}}
\subfigure{\includegraphics[scale=1]{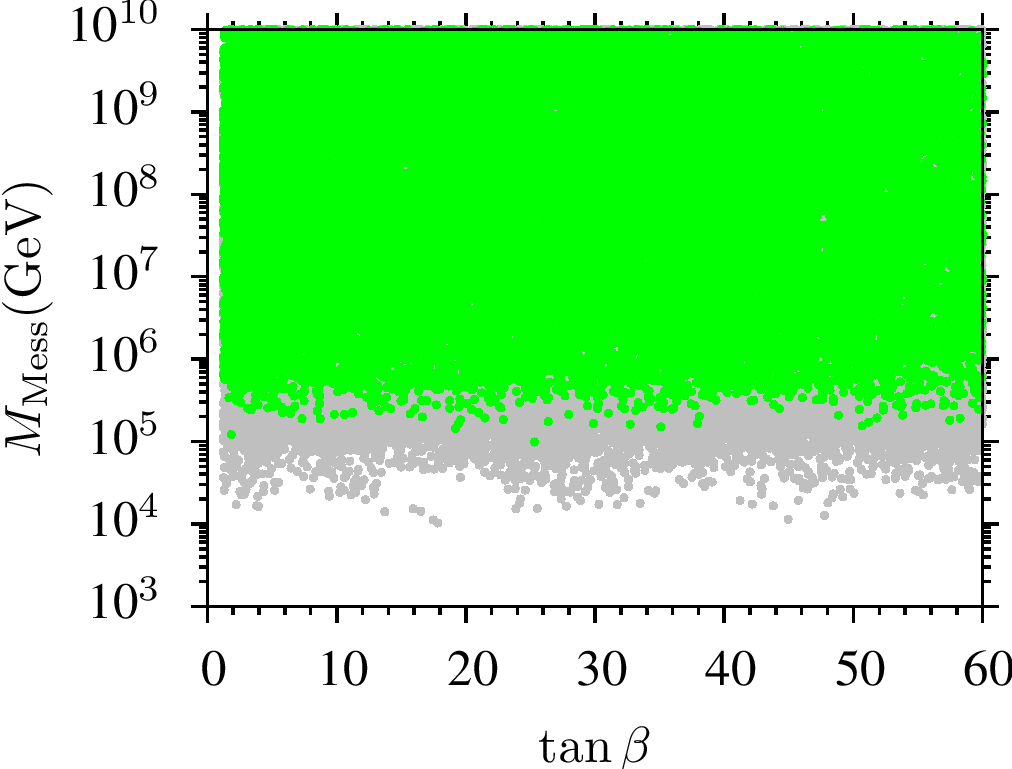}}
\subfigure{\includegraphics[scale=1]{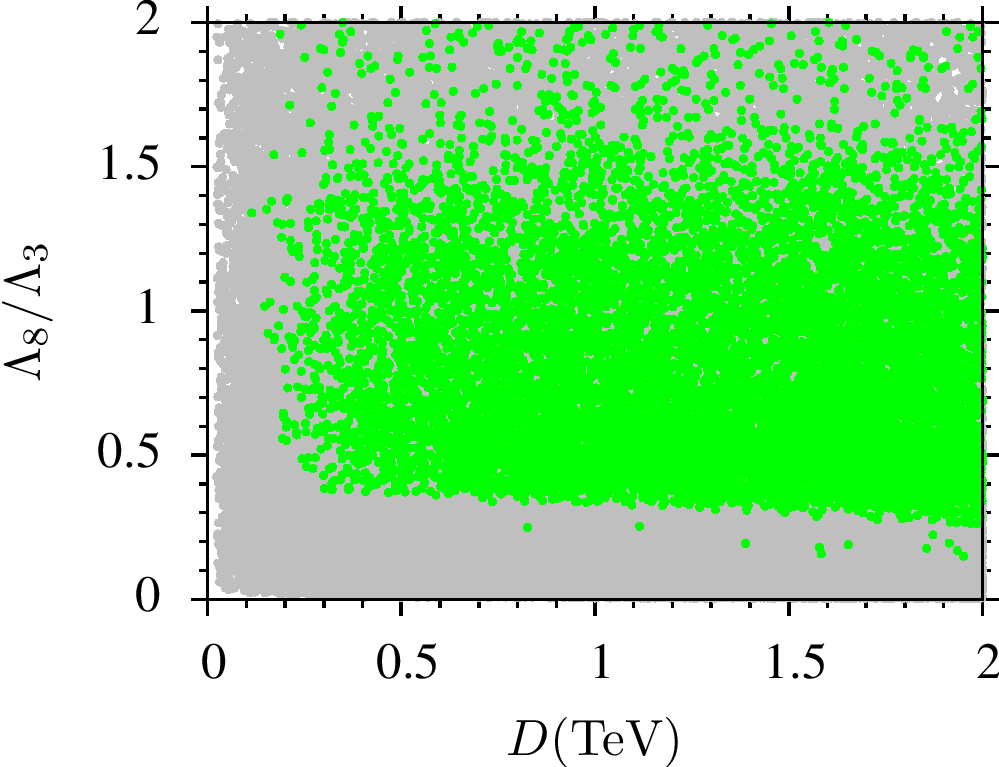}}
\subfigure{\hspace{2.0cm}\includegraphics[scale=1]{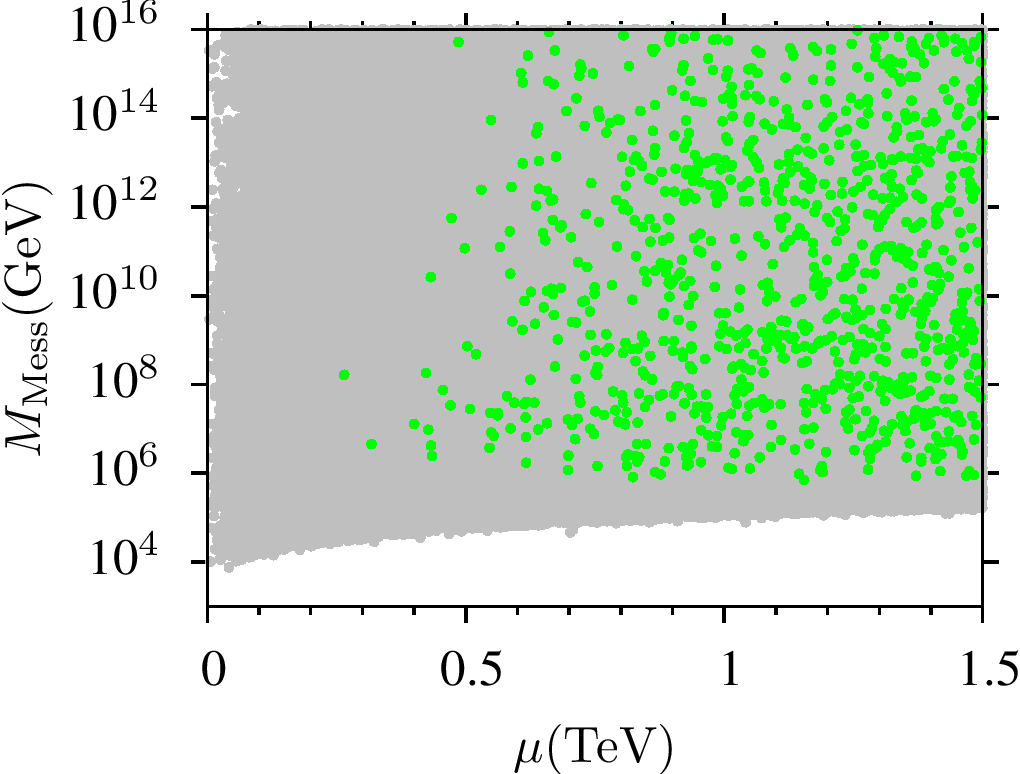}}
\caption{Plots in the $M_{{\rm Mess}}-\Lambda_{3}$, $M_{{\rm Mess}}-\Lambda_{8}$, $M_{{\rm Mess}}-\tan\beta$, $\Lambda_{8}-\Lambda_{3}$, $\Lambda_{8}/\Lambda_{3}-D$, and $\mu-D$ planes.
All points are consistent with REWSB. Green points are consistent with the experimental constraints in Table~\ref{table1}.}
\label{fig1}
\end{figure}
%
%
%
%
\begin{figure}[t!]
\subfigure{\includegraphics[scale=1]{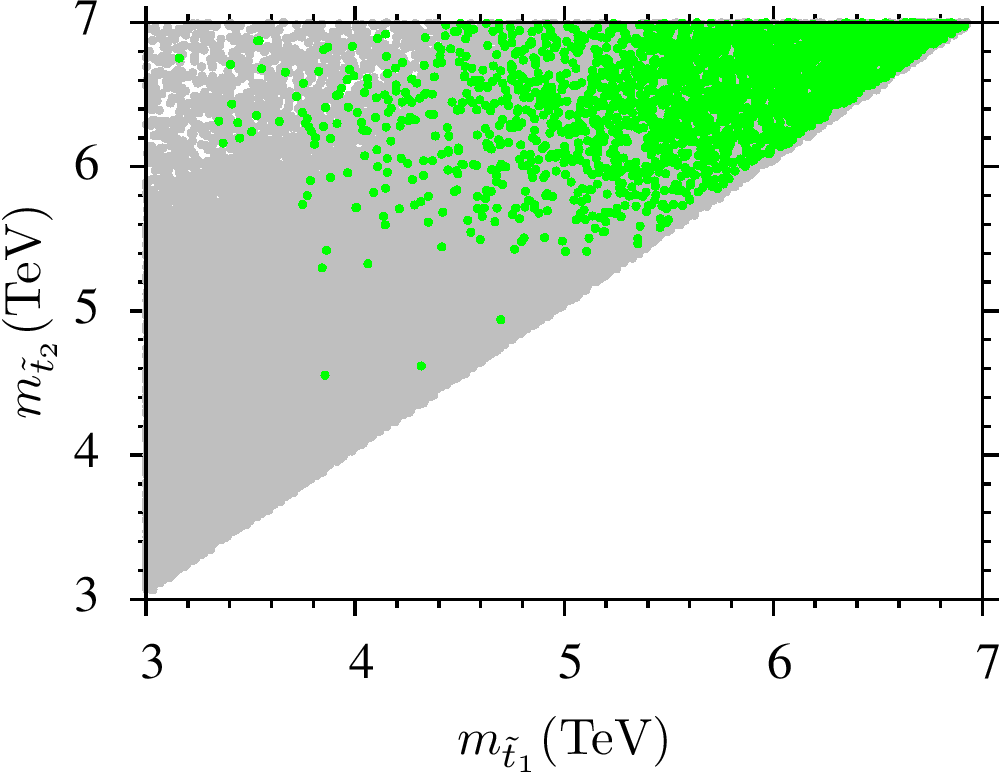}}
\subfigure{\hspace{2.1cm}\includegraphics[scale=1]{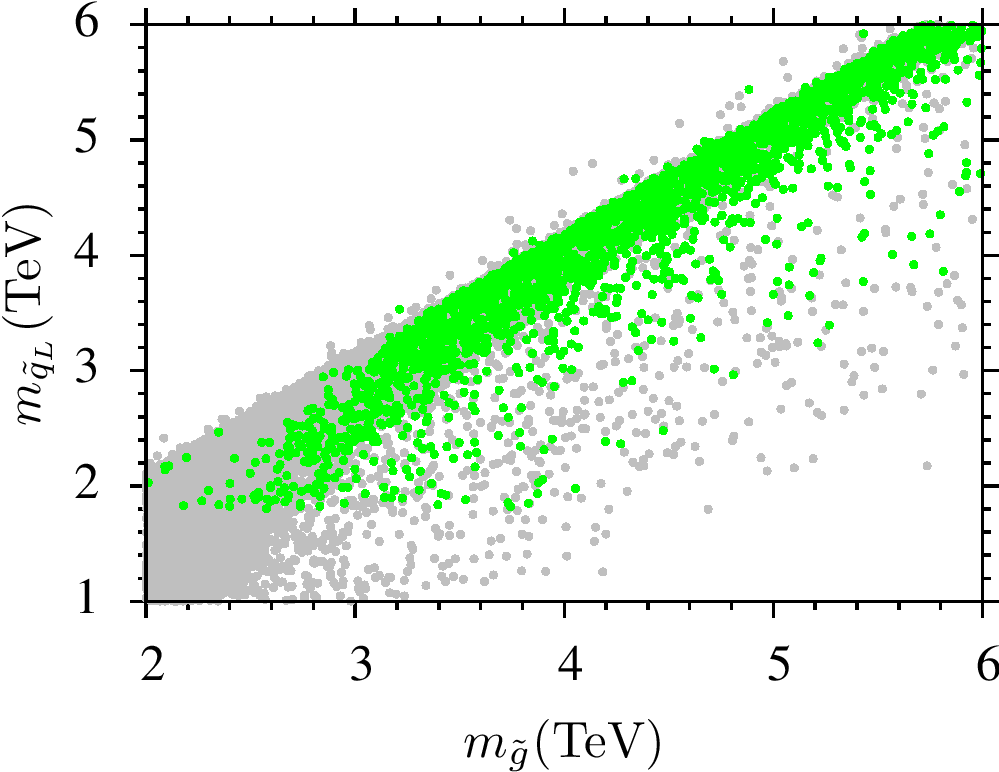}}
\subfigure{\includegraphics[scale=1]{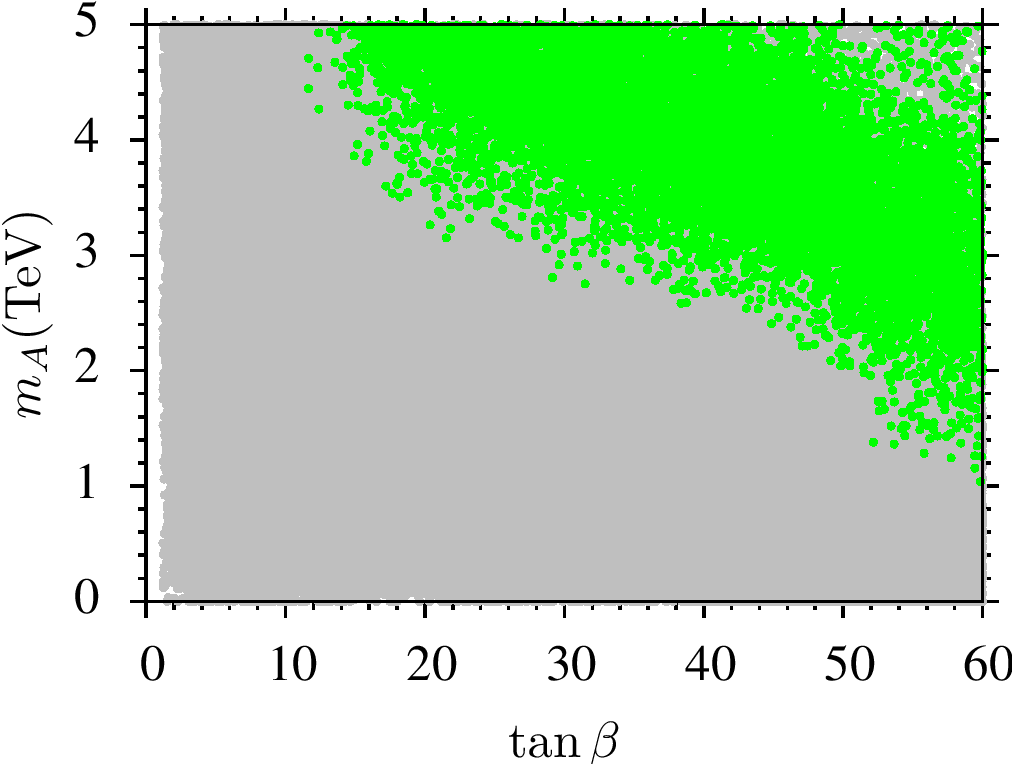}}
\subfigure{\hspace{2.1cm}\includegraphics[scale=1]{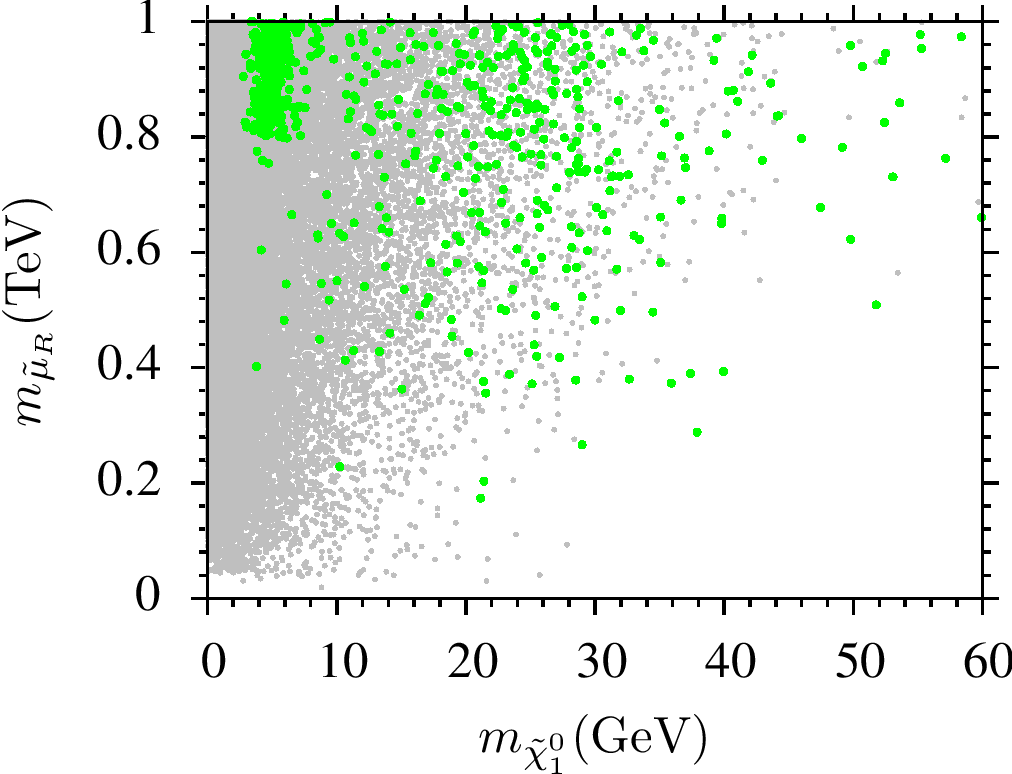}}
\caption{Plots in the   $m_{\tilde{\tau}_{2}}-m_{\tilde{\tau}_{1}}$, $m_{A}-\tan\beta$, $m_{\tilde{\mu}_{R}}-m_{\tilde{\chi}_{1}^{0}}$ and
$m_{\tilde{\mu}_{L}}-m_{\tilde{\chi}_{1}^{0}}$ planes. The color coding is the same as in Figure~\ref{fig1}.}
\label{fig2}
\end{figure}
\begin{figure}[t!]
\centering
\subfigure{\includegraphics[scale=1.]{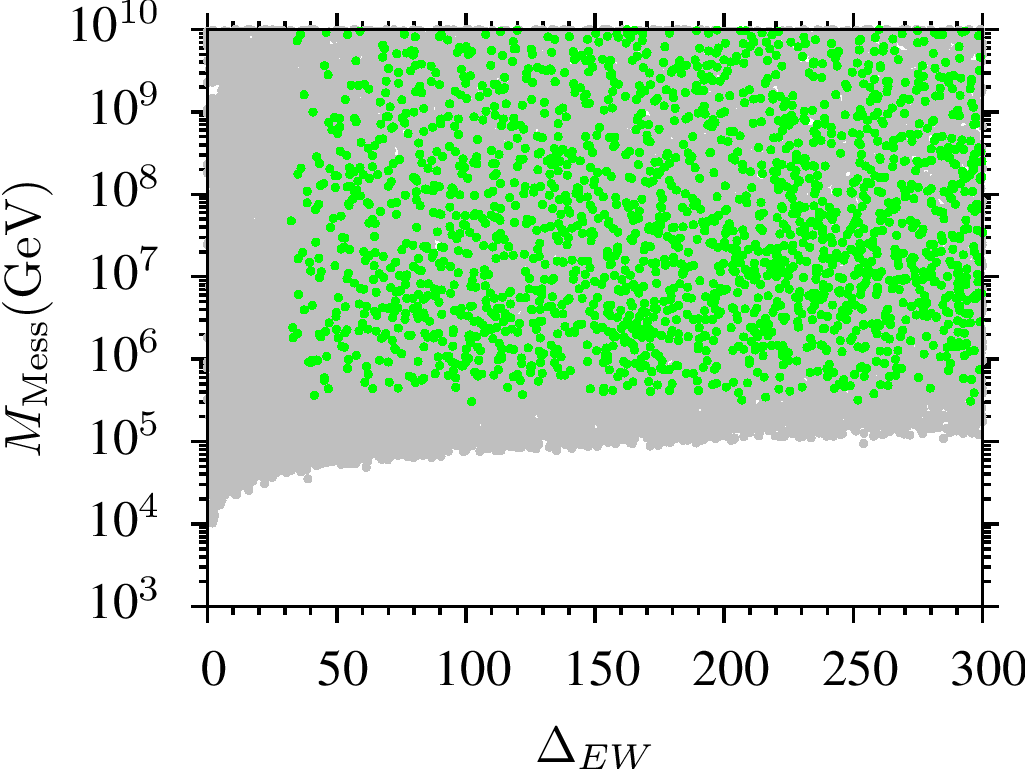}}
\subfigure{\hspace{1.3cm} \includegraphics[scale=1.]{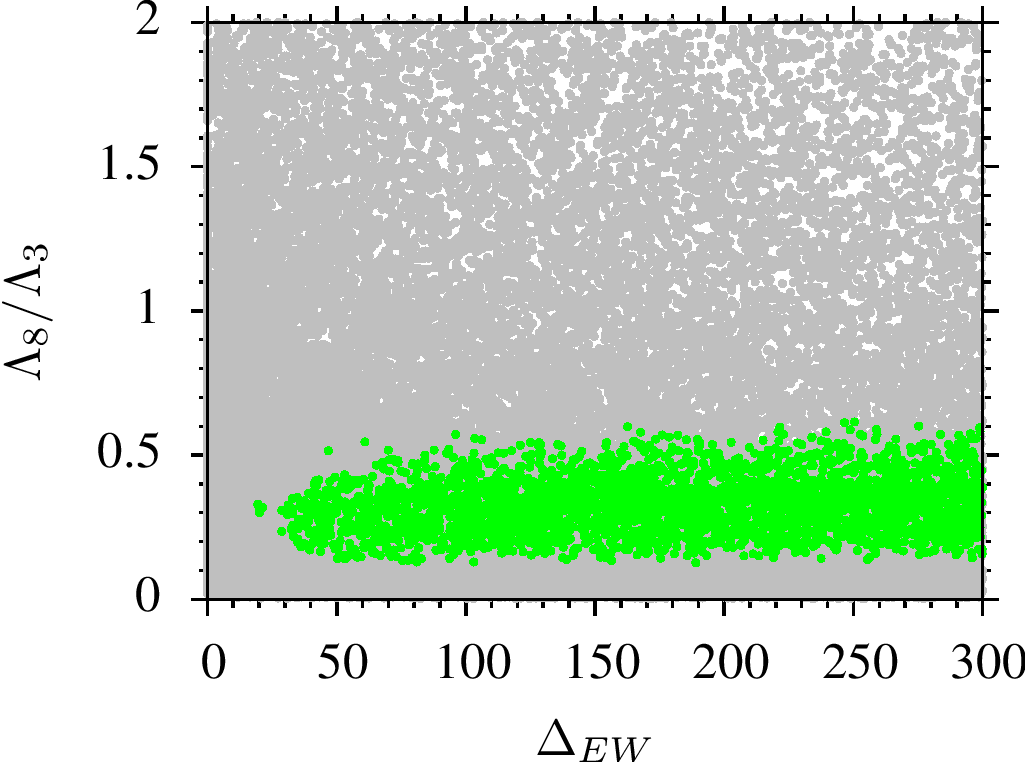}}
\caption{Plots in the $M_{{\rm Mess}}-\Delta_{EW}$ and $\Lambda_{8}/\Lambda_{3}-\Delta_{EW}$  planes. The color coding is the same as in Figure~\ref{fig1}.}
\label{fig3}
\end{figure}
\begin{figure}[t!]
\centering
\subfigure{\includegraphics[scale=1.5]{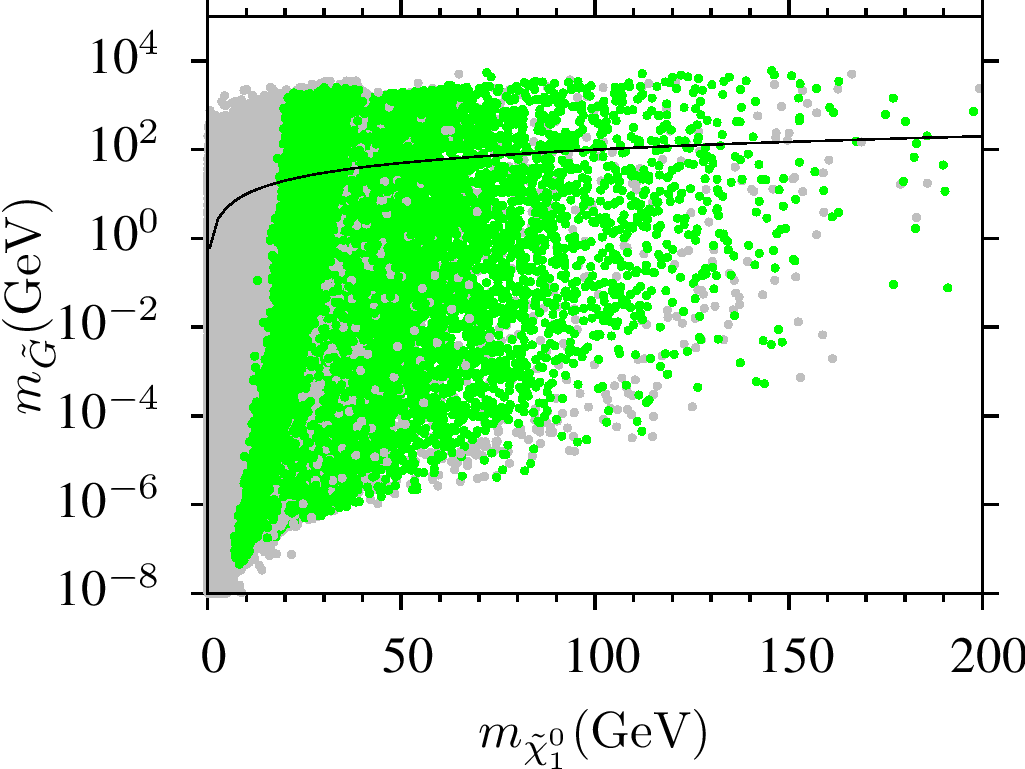}}
\caption{Gravitino and neutralino masses in the $m_{\tilde{G}}-m_{\tilde{\chi}_{1}^{0}}$ plane. The color coding is the same as in Figure~\ref{fig1}.
The solid line indicates regions where $m_{\tilde{G}}=m_{\tilde{\chi}_{1}^{0}}$.}
\label{fig4}
\end{figure}

In this section we present the results of the scan over the parameter space listed in Eq.~(\ref{parameterRange}).  As previously mentioned,
the characteristic feature of our model  is that all  sfermions receive an additional  $U(1)_{B-L}$ $D$-term  contribution to their SSB masses,
see Eq.~(\ref{sGMSB_masses3}).
Figure~\ref{fig1} represents plots in  $M_{{\rm Mess}}-\Lambda_{3}$, $M_{{\rm Mess}}-\Lambda_{8}$,  $\Lambda_{8}-\Lambda_{3}$, $M_{{\rm Mess}}-\tan\beta$, $\Lambda_{8}/\Lambda_{3}-D$, and $\mu-D$ planes with all points being consistent with REWSB. Green points are consistent with the experimental constraints presented in Table~\ref{table1}. We see from the
$M_{{\rm Mess}}-\Lambda_{3}$ plane that  $M_{{\rm Mess}}$ and $\Lambda_{3}$ parameters can lie in a wide range consistent
with the current experimental constraints. It is interesting to note that $\Lambda_{3}$ can even be as low as $2\times 10^{4}$~GeV.
On the other hand, from the $M_{{\rm Mess}}-\Lambda_{8}$ plane, $\Lambda_{8}$ is bounded at about $10^{5}$~GeV from below.
The $M_{{\rm Mess}}-\tan\beta$ panel indicates that there is a slight preference for $\tan\beta>10$ when $M_{{\rm Mess}}<10^6$~GeV.

The $\Lambda_{8}-\Lambda_{3}$ and $\Lambda_{8}/ \Lambda_{3}-D$ planes show that $\Lambda_{8}$ should be larger than $\Lambda_{3}$
over most of the parameter space, even though it is possible to have $\Lambda_{8} \leq \Lambda_{3}$ in a small portion of the parameter space,
particularly when $\Lambda_3>10^5$~GeV.
The D-term contribution to the scalars can be as low as about 100~GeV,  which means that experimentally acceptable ($O(100 \rm~GeV)$) slepton masses
cannot be generated through RGE evolution if their masses are negligibly  small at the messenger scale.
Finally, the $\mu-D$ panel shows that the MSSM $\mu -$term as low as 200-300 GeV can be realized. This indicates that in this model
the little hierarchy problem is not as severe as in the minimal GMSB scenario with messenger fields allocated in $5+\bar5$  (or $10+\overline{10}$)
representations of SU(5). A more detailed analysis about the little hierarchy problem will be presented when we discuss  Figure 3.

Figure~\ref{fig2} displays the mass spectrum with plots in the $m_{\tilde{t}_{2}}-m_{\tilde{t}_{1}}$, $m_{\tilde{q}}-m_{\tilde{g}}$,
$m_{A}-\tan\beta$  and $m_{\tilde{\mu}_{R}}-m_{\tilde{\chi}_{1}^{0}}$ planes. The color coding is the same as Figure~\ref{fig1}.
The colored sparticles are rather heavy as seen from the top panels of Figure~\ref{fig2}, but still there is hope that some of them can be tested at the LHC.
The stops  are required to be heavier than about 4~TeV in order to satisfy the Higgs mass bound.
The gluinos and squarks from the first two families ($\tilde{q}_{L}$) can be as light as  2~TeV or so.
The $m_{A}-\tan\beta$ plane reveals the correlation between the mass of the CP-odd $A-$boson and $\tan\beta$.
We see that $m_{A} \sim 1$~TeV is achieved for $\tan\beta \approx 60$. For small  and moderate $\tan\beta$ values, $m_{A} $ is more than 2~TeV
and it would be difficult to detect A-boson at the LHC.

Finally the $m_{\tilde{\mu}_{R}}-m_{\tilde{\chi}_{1}^{0}}$ plot shows that in our scenario it is possible to have light sleptons.
In particular, we see that the right-handed smuon can be around 200~GeV, which makes it accessible at the LHC.
Having such a light smuon in the spectrum can be helpful, in principle, for the muon $g-2$ anomaly~\cite{BNL}.
However, in our scenario the bino mass parameter $M_{1}$ has negative values as a result of its RGE evolution,
which gives a sizable contribution to muon $g-2$ but with the wrong sign.
This explains why we cannot provide resolution of the muon $g-2$ anomaly from SUSY contributions in our scenario.

The latest (7.84) version of  ISAJET~\cite{Paige:2003mg} calculates the  fine-tuning conditions related to the little hierarchy problem at $M_{EW}$.
Including the one-loop effective potential contributions to the tree level MSSM Higgs potential, the Z boson mass is given by the  relation:
\beq
\frac{M_Z^2}{2} =
\frac{(m_{H_d}^2+\Sigma_d^d)-(m_{H_u}^2+\Sigma_u^u)\tan^2\beta}{\tan^2\beta
-1} -\mu^2 \; .
\label{eq:mssmmu}
\eeq
The $\Sigma$'s stand for the contributions arising from the one-loop effective potential (for more details see ref.~\cite{Baer:2012mv}).
All parameters  in Eq.~(\ref{eq:mssmmu}) are defined at the weak scale $M_{EW}$.

In order to measure the EW scale fine-tuning condition associated with the little hierarchy problem, the following definitions are used~\cite{Baer:2012mv}:
\beq
 C_{H_d}\equiv |m_{H_d}^2/(\tan^2\beta -1)|,\,\, C_{H_u}\equiv
|-m_{H_u}^2\tan^2\beta /(\tan^2\beta -1)|, \, \, C_\mu\equiv |-\mu^2 |,
\label{cc1}
\eeq
 with each $C_{\Sigma_{u,d}^{u,d} (i)}$  less than some characteristic value of order $M_Z^2$.
Here, $i$ labels the SM and supersymmetric particles that contribute to the one-loop Higgs potential.
For the fine-tuning condition we have
\beq
 \Delta_{\rm EW}\equiv {\rm max}(C_i )/(M_Z^2/2).
\label{eq:ewft}
\eeq
Note that Eq.~(\ref{eq:ewft}) defines the fine-tuning  condition at $M_{EW}$ without addressing
the question of the origin of the parameters that are involved. Hence, $\Delta_{\rm EW}$ represents a lower bound on fine-tuning~\cite{DEW}.

As mentioned earlier, the little hierarchy problem can be ameliorated in our model, since $M_{3}$ and $M_{2}$ are generated by
the two free parameters $\Lambda_{8}$ and $\Lambda_{3}$ respectively. It can be quantified with $\Delta_{EW}$ as shown in Figure~\ref{fig3}
with plots in the $M_{{\rm Mess}}-\Delta_{EW}$ and $\Lambda_{8}/ \Lambda_{3} -\Delta_{EW}$  planes.
The color coding is the same as Figure~\ref{fig1}. Acceptable fine-tuning is usually assumed when $\Delta_{EW} \leq 10^{2}$,
and our model can yield $\Delta_{EW} \sim 30$ or so.  As seen from the $M_{{\rm Mess}}-\Delta_{EW}$ plane,
it is possible to have solutions with $\Delta_{EW} \leq 100$ even for high values of $M_{\mathrm{Mess}}<10^{10}$~GeV.
It is interesting to note that the solution with relatively small $\Delta_{EW}$ appears with $0.1<\Lambda_{8}/ \Lambda_{3}< 0.5$,
which is a clear indication of the necessity of non-universal gauginos in the spectrum. A similar observation in gravity mediation scenario was made sometime ago
in ref.~\cite{Gogoladze:2009bd}.

Figure~\ref{fig4} presents results for the gravitino and neutralino masses in the $m_{\tilde{G}}-m_{\tilde{\chi}_{1}^{0}}$ plane.
The color coding is the same as Figure~\ref{fig1}. The solid line indicates regions where $m_{\tilde{G}}=m_{\tilde{\chi}_{1}^{0}}$.
The LSP in gauge mediation models is usually the gravitino since its mass is expected to be much smaller than the typical sparticle mass.  In our model the gravitino mass varies in a wide range
from eV to TeV scales, and it is found to be the LSP over most of the parameter space. In standard scenarios,
the WMAP (and Planck) bound on the LSP relic density ($\Omega h^{2} \simeq 0.11$~\cite{Hinshaw:2012aka}) yields gravitino mass $\sim 200$~eV,
which makes the gravitino a candidate for hot dark matter. However, the latter cannot contribute more than 15\% to the dark matter density and this, in turn, requires the gravitino mass to be less than 30~eV~\cite{Viel:2005qj}. In this context, the gravitino can manifest itself through missing energy in colliders.  In order
to have a complete dark matter scenario one could invoke axions as cold dark matter in
this region.

A gravitino mass $\gtrsim30$~eV requires non-standard scenarios in order to agree with observations.  
Such non-standard scenarios include gravitino decoupling and freezing out earlier than in the standard scenario,  
which  may be possible in a theory  with  more degrees of freedom than the MSSM~\cite{Feng:2010ij}.  
A gravitino of mass $ \gtrsim$~keV  is still possible and it can be cold enough to constitute all 
of the dark matter if non-standard scenarios, such as early decoupling, is assumed.

Finally we display three benchmark points which exemplify our findings, with all masses in GeV units. Point 1 exemplifies a solution with a LSP neutralino with a mass of about 80~GeV,
even though the other sparticles are rather heavy. This point shows that regions of the parameter space which yield LSP neutralino,
can be realized for a high $M_{\mathrm{Mess}} \sim 10^{15}$~GeV, and  require rather  high fine-tuning ($\Delta_{EW}\sim 7600$).
Point 2 represents a solution with a LSP gravitino with mass $\sim$~keV scale. Such solutions can be obtained for low
$M_{\mathrm{Mess}}$ values ($\sim 10^{7}$~GeV) with moderate to low fine-tuning. Similarly, Point 3 displays a solution with LSP gravitino of mass $\sim$ 27~eV.
The messenger scale and the required fine-tuning are low, similar to Point 2. The solutions exemplified by Point 3 offer gravitino
as a plausible hot dark matter candidate.

\begin{table}[ht!]
\centering
\scalebox{1.0}{
\begin{tabular}{|l|ccc|}
\hline
                 & Point 1 & Point 2 & Point 3  \\
\hline
\hline
$\Lambda_{3}$  & $0.12\times 10^{4}$  & $0.49\times 10^{6}$ & $0.61\times 10^{5}$ \\
$\Lambda_{8}$  & $0.15\times 10^{7}$  & $0.14\times 10^{6}$ & $0.28\times 10^{6}$ \\
$M_{\rm mess}$ & $0.52\times 10^{16}$ & $0.22\times 10^{7}$ & $0.54\times 10^{6}$ \\
$\tan\beta$    & 54.4 & 34.9 & 57.2 \\
$D$            & 1455 & 1578 & 1914 \\
\hline
$\mu$          & 6012 & 487 & 3086 \\
$\Delta_{EW}$  & 2920 & 54 & 1946 \\
\hline
$m_h$           & 125.4 & 124.5 & 124.4 \\
$m_H$           & 6003  & 1935  & 1867 \\
$m_A$           & 5992 & 1923  & 1855 \\
$m_{H^{\pm}}$   & 6004 & 1937 & 1869 \\

\hline
$m_{\tilde{\chi}^0_{1,2}}$
                 & \textbf{81.9}, 223  & 5.01, 480 & 6.1, 329  \\

$m_{\tilde{\chi}^0_{3,4}}$
                 & 5113, 5113  & 482, 2593 & 2796, 2796 \\

$m_{\tilde{\chi}^{\pm}_{1,2}}$
                & 223, 5365 & 493, 2552 & 330, 2757 \\

$m_{\tilde{g}}$  & 9418 & 3115 & 5745 \\
\hline $m_{ \tilde{u}_{L,R}}$
                 & 8324, 8361  & 4058, 5450  & 7719, 5576 \\
$m_{\tilde{t}_{1,2}}$
                 & 5017, 5128 & 3679, 5038 & 5159, 7343 \\
\hline $m_{ \tilde{d}_{L,R}}$
                 & 8322, 8334  & 4059, 3303  & 5598, 5576 \\
$m_{\tilde{b}_{1,2}}$
                 & 7119, 7240  & 3213, 3747 & 5251, 5470 \\
\hline
$m_{\tilde{\nu}_{e,\mu}}$
                 & 3886 & 5270 & 5708 \\
$m_{\tilde{\nu}_{\tau}}$
                 & 3714 & 5217  & 5594 \\
\hline
$m_{ \tilde{\mu}_{L,R}}$
                & 1942, 2809  & 5272, 4871  & 5863, 5607 \\
$m_{\tilde{\tau}_{1,2}}$
                &1652, 2202  & 4753, 5213  & 5547, 5633 \\
\hline
$m_{\tilde{G}}$  & $1833$  & $2.7 \times 10^{-7}$ & $36 \times 10^{-9}$\\
\hline
\end{tabular}}
\caption{Benchmark points for exemplifying our results. All masses are in GeV. Point 1 exemplifies a solution for neutralino LSP.
Point 2 depicts a solution with a low $\Delta_{EW}$. Finally, Point 3 represents gravitino as hot dark matter solution.}
\label{benchsgmsb}
\end{table}


\section{Conclusion}
\label{sec:conclusion}
We have explored the spectroscopy and related topics in a class of models within the framework of gauge mediation supersymmetry
breaking where the messenger fields transform in the adjoint representation of the Standard Model gauge symmetry.
To avoid ``massless" or too light right-handed sleptons a  non zero $U(1)_{B-L}$ D-term is introduced. This  provides
additional  contributions  to  the  soft  supersymmetry  breaking  mass  terms and makes the right-handed slepton masses compatible with the current experimental data.
In this framework we show that the observed 125~GeV Higgs boson mass and the desired relic dark matter abundance  can be simultaneously accommodated
with relatively light sleptons accessible at the LHC. In the spectrum we do have relatively light smuons but due to the negative sign of bino mass
at low scale the supersymmetric contribution either comes with the wrong sign or is not significant enough to explain the muon $g-2$ anomaly.

\section*{Acknowledgments}
 We thank Bhaskar Dutta  and Tianjun Li  for helpful discussions. This work is supported in part by the DOE Grant DE-SC0013880  (Q.S.),   Bartol Research Institute (I.G.), the Rustaveli National Science Foundation No. 03/79 (I.G.), and The Scientific and Technological Research Council of Turkey (TUBITAK)
Grant no. MFAG-114F461 (CS\"{U}). This work used the Extreme Science and Engineering Discovery Environment (XSEDE),
which is supported by the National Science Foundation grant number OCI-1053575.  Part of the numerical calculations reported in this paper
were performed at the National Academic Network and Information Center (ULAKBIM) of TUBITAK, High Performance and Grid Computing Center (TRUBA Resources).


\end{document}